\documentclass[sigconf]{acmart}
\usepackage{algorithm}
\usepackage{algorithmic}

\usepackage{multirow}
\usepackage{soul}

\AtBeginDocument{%
  \providecommand\BibTeX{{%
    \normalfont B\kern-0.5em{\scshape i\kern-0.25em b}\kern-0.8em\TeX}}}

\copyrightyear{2023}
\acmYear{2023}
\setcopyright{acmlicensed}\acmConference[WWW '23]{Proceedings of the ACM Web Conference 2023}{May 1--5, 2023}{Austin, TX, USA}
\acmBooktitle{Proceedings of the ACM Web Conference 2023 (WWW '23), May 1--5, 2023, Austin, TX, USA}
\acmPrice{15.00}
\acmDOI{10.1145/3543507.3583240}
\acmISBN{978-1-4503-9416-1/23/04}

\begin{document}

%%
%% The "title" command has an optional parameter,
%% allowing the author to define a "short title" to be used in page headers.
\title{Fine-tuning Partition-aware Item Similarities for Efficient and Scalable Recommendation}

%%
%% The "author" command and its associated commands are used to define
%% the authors and their affiliations.
%% Of note is the shared affiliation of the first two authors, and the
%% "authornote" and "authornotemark" commands
%% used to denote shared contribution to the research.
\author{Tianjun Wei}
\email{tjwei2-c@my.cityu.edu.hk}
\orcid{0000-0001-7311-7101}
\affiliation{%
  \institution{City University of Hong Kong}
  \city{Kowloon}
  \country{Hong Kong}
}

\author{Jianghong Ma}
\authornote{Corresponding author.}
\email{majianghong@hit.edu.cn}
\orcid{0000-0002-0524-3584}
\affiliation{%
 \institution{Harbin Institute of Technology}
 \city{Shenzhen}
 \country{China}}

\author{Tommy W. S. Chow}
\email{eetchow@cityu.edu.hk}
\orcid{0000-0001-7051-0434}
\affiliation{%
  \institution{City University of Hong Kong}
  \city{Kowloon}
  \country{Hong Kong}
}

%%
%% By default, the full list of authors will be used in the page
%% headers. Often, this list is too long, and will overlap
%% other information printed in the page headers. This command allows
%% the author to define a more concise list
%% of authors' names for this purpose.
% \renewcommand{\shortauthors}{Trovato and Tobin, et al.}

%%
%% The abstract is a short summary of the work to be presented in the
%% article.
\begin{abstract}
    Collaborative filtering (CF) is widely searched in recommendation with various types of solutions. Recent success of Graph Convolution Networks (GCN) in CF demonstrates the effectiveness of modeling high-order relationships through graphs, while repetitive graph convolution and iterative batch optimization limit their efficiency. Instead, item similarity models attempt to construct direct relationships through efficient interaction encoding. Despite their great performance, the growing item numbers result in quadratic growth in similarity modeling process, posing critical scalability problems. In this paper, we investigate the graph sampling strategy adopted in latest GCN model for efficiency improving, and identify the potential item group structure in the sampled graph. Based on this, we propose a novel item similarity model which introduces graph partitioning to restrict the item similarity modeling within each partition. Specifically, we show that the spectral information of the original graph is well in preserving global-level information. Then, it is added to fine-tune local item similarities with a new data augmentation strategy acted as partition-aware prior knowledge, jointly to cope with the information loss brought by partitioning. Experiments carried out on 4 datasets show that the proposed model outperforms state-of-the-art GCN models with 10x speed-up and item similarity models with 95\% parameter storage savings.
\end{abstract}

%%
%% The code below is generated by the tool at http://dl.acm.org/ccs.cfm.
%% Please copy and paste the code instead of the example below.
%%


%%
%% Keywords. The author(s) should pick words that accurately describe
%% the work being presented. Separate the keywords with commas.

%% A "teaser" image appears between the author and affiliation
%% information and the body of the document, and typically spans the
%% page.
% \begin{teaserfigure}
%   \includegraphics[width=\textwidth]{sampleteaser}
%   \caption{Seattle Mariners at Spring Training, 2010.}
%   \Description{Enjoying the baseball game from the third-base
%   seats. Ichiro Suzuki preparing to bat.}
%   \label{fig:teaser}
% \end{teaserfigure}

%%
%% This command processes the author and affiliation and title
%% information and builds the first part of the formatted document.
\maketitle

\section{Introduction}
The rapid development of the Internet has given rise to recommender systems which focus on alleviating the information overload problem by providing personalized recommendations. The core task of recommendation is to capture user preferences through the interactions between the users and the recommended objects, i.e., the items. This task, known as collaborative filtering (CF) \cite{He2017, Xie2021}, has been extensively studied in recent years by academia and industry.

A straightforward approach to achieve CF is to model the relationships for each pair of items. Historically, several studies \cite{Deshpande2004, Sarwar2001} attempt to estimate item relationships through adopting metrics like Cosine similarity. These non-parametric heuristic models are simple and efficient, while yielding inferior recommendation performance \cite{Christakopoulou2016}. Subsequent studies propose item similarity models to solve an encoding problem of the user-item interaction matrix \cite{Ning2011, Steck2019}. Parameters learned in the encoding problem exhibit direct relationship mapping across items, enabling the  generation of good-quality recommendations. Despite their effectiveness, problems are encountered when the quantity of items grows rapidly, which is occurring in today's Internet applications. The effort to model item similarities follows a quadratic growth with an increasing scale of item sets, which yields scalability problems in production environments.

Since each entry in the user-item interaction matrix in CF can be naturally considered as an edge between nodes in a bipartite graph, graph convolutional networks (GCN) models \cite{Wang2019, He2020} are proposed to explore high-order user-item relationships. Through stacking multiple convolution layers of interaction bipartite graph, GCN models can capture multi-hop relationships across users and items. However, too many layers lead to high computational cost and over-smooth problems, which are argued by recent studies \cite{Shen2021, Mao2021}. In a recent study \cite{Mao2021}, UltraGCN is proposed to approximate infinite layer linear GCN through 1-hop user-item and 2-hop item-item relationships, demonstrating superior performance over existing GCN models. As the 2-hop item-item graph becomes much denser compared to user-item graph, UltraGCN introduces graph sampling to keep the most informative item-item pairs. Satisfactory results in UltraGCN indicate that valid information is retained by sampled item-item graph, which motivates us to further explore how it works.

To answer the question, we conduct analysis on the sampled item-item graph in UltraGCN. The investigation shows that as the number of neighbors retained in the sampled graph decreases, the algebraic connectivity of the graph gradually decreases and even becomes 0. This observation indicates that the items are divided into disconnected groups and suggests that the graph sampling strategy has the potential to detect partition structures in item-item graph. To obtain a complete understanding of this phenomenon, we analyze the derivation of the sampling strategy in UltraGCN, and show that it is a local approximation of modularity maximization \cite{Newman2004}, which is commonly adopted in graph community detection \cite{Shi2000, Newman2013}. It naturally raises the assumption that latent partitions are located in the item set, which can be detected to benefit the item relationship modeling. When considering the reification of items in recommender systems, which typically refer to products, services, and medias with diverse characteristics, this assumption can also be justified. It provides an inspiration to identify closely related items through partitioning item-item graph to reduce the effort in similarity modeling, thus enhance the scalability of item similarity models. However, trivially adopting similarity modeling within each partition may suffer from information loss and show a trade-off between efficiency and accuracy, as neither the graph partitioning results nor the real-world training data are ideal \cite{Khawar2020}. Therefore, it is still challenging to leverage partition information for efficient and scalable item similarity modeling, while maintaining the robustness of the recommender system.

In this paper, we seek approaches to effectively leverage information in the partitioned item-item graph to address the above challenge, and finally propose a novel model called Fine-tuning Partition-aware item Similarities for efficient and scalable Recommendation (FPSR). Specifically, we investigate the spectral partitioning strategy adopted on the item-item graph, and show that the eigenvectors of graph Laplacian derived for graph partitioning are powerful in preserving inter-partition item relationships. Therefore, we include the global-level spectral information in the item similarity modeling process within each partition, allowing FPSR to maintain well recommendation capability on the entire item set with significant reduction on computational costs. Moreover, we propose a data augmentation approach to explicitly add the partition information in the encoding problem of the interaction matrix, which acts as the prior knowledge of the partition-aware similarity fine-tuning. Extensive experiments conducted on 4 real-world public datasets demonstrate better performance, efficiency and scalability of FPSR, with remarkable savings of 90\% training time compared to state-of-the-art GCN models and 95\% parameter storage compared to latest item similarity models on all datasets. The PyTorch implementation of our proposed FPSR model is available at: \url{https://github.com/Joinn99/FPSR}.

To summarize, the contributions in this work are listed below:
\begin{itemize}
    \item We empirically and experimentally analyze the features and connections of the graph sampling strategy in UltraGCN and graph partitioning algorithms, and find a clear path to achieve scalable item similarity modeling by item-item graph partitioning.
    \item We propose a novel item similarity model for recommendation named FPSR, which performs fine-tuning of the intra-partition similarities with the leveraging of global-level information across the entire item set.
    \item We conduct a comprehensive experimental study on 4 real-world datasets to show the significant advantages of FPSR in terms of accuracy, efficiency,  and scalability compared to the latest GCN models and item similarity models.
\end{itemize}

\section{Investigation Of Graph-based CF}
\subsection{Problem Formulation}
Suppose a user set $\mathcal{U}$ , an item set $\mathcal{I}$, and the observed interaction set $\mathcal{T}=\{(u, i) \lvert u \in \mathcal{U}, i \in \mathcal{I}\}$ between users and items. For each user $u$, Collaborative Filtering (CF) aims to recommend top-$K$ items from $\mathcal{I}$ this user has not interacted with.

\subsection{Graph Construction in CF}
\label{Subsec:GraphCF}
In CF task, user preferences are modeled based on their interactions with items. Here, we adopt the settings in existing methods \cite{Ning2011, Wang2019} to define the user-item interaction matrix $R\in\{0,1\}^{\lvert \mathcal{U} \lvert \times \lvert \mathcal{I} \lvert}$ with implicit feedbacks as
\begin{equation}
R_{ui}= 
\begin{cases}
    1, & \text{if interaction }(u, i)\text{ is observed,}\\
    0, & \text{otherwise.}
\end{cases}
\end{equation}
As a common practice, standard normalization is applied to ensure the stability of multi-layer graph convolution. The normalized interaction matrix is defined in \cite{Kipf2017} as follows:
\begin{equation}
\tilde{R}=D_U^{-\frac{1}{2}}RD_I^{-\frac{1}{2}},
\end{equation}
where $D_I=diag(\mathbf{1}^TR)$ is the row sum of the interaction matrix $R$. Similarly, the column sum of $R$ is defined as $D_U=diag(R\mathbf{1})$. On the other hand, the item-item adjacency matrix is also explored to identify the relationships between items. The normalized item-item adjacency matrix can be constructed as \cite{Shen2021}:
\begin{equation}
\tilde{Q}=\tilde{R}^T\tilde{R}.
\label{Eq:NormQ}
\end{equation}
And the unnormalized adjacency matrix for items $Q$ can be derived similarly.

\subsection{Discover Group Structure in Item-item Graph}
\begin{figure}[t!]
  \centering
  \includegraphics[width=2.4in]{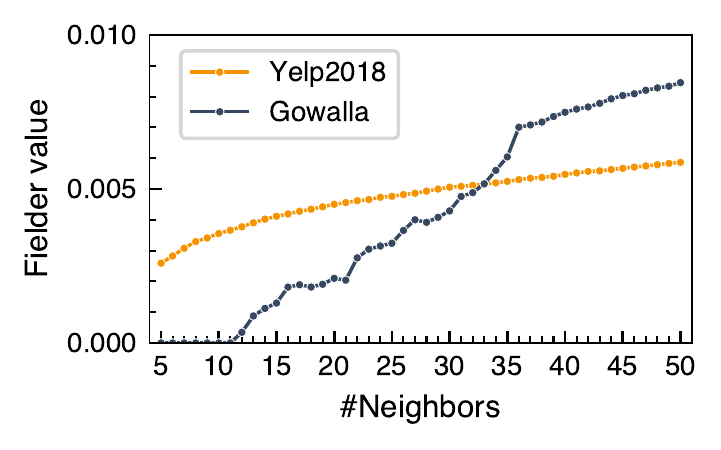}
  \caption{Algebraic connectivity (Fielder value) of the sampled item-item graph in UltraGCN.}
  \Description{Line graph showing the algebraic connectivity from 0 to 0.01 on the Y axis against of the number of neighbors in sampled item-item graph in UltraGCN from 5 to 50 on the X axis. Two lines are shown. When the number of neighbors decreases from 50 to 5, Yelp2018 shows a slow decrease from 0.006 to 0.004, and Gowalla shows a faster decline from 0.009 and reaches 0 at the number of neighbors with 11.}
  \label{Fig:Fielder}
  \end{figure}

Since the item relationships in item-item graph are essentially the second-order relations in the user-item interaction graph, the item-item adjacency matrix $Q$ is usually much denser than $R$. Therefore, directly adopting $Q$ in the model training will substantially increase the time and storage complexity. To retain high training efficiency, UltraGCN \cite{Mao2021} performs sampling in $Q$ to keep the most informative item-item pairs. Top-$k$ item-item pairs in the row $i$ of $Q$ are selected according to
\begin{equation}
\mathop{\arg\max}\limits_{j\in \mathcal{I}}^k\ \omega_{ij}.
\label{Eq:GP_UltraGCN_Filtering}
\end{equation}
Here the weight $\omega_{ij}$ is defined as
\begin{equation}
\omega_{ij}=\frac{Q_{ij}}{\sqrt{p_j}}\cdot\frac{\sqrt{p_i}}{p_i-Q_{ii}},
\label{Eq:GP_UltraGCN_Omega}
\end{equation}
where $p_i=\sum_kQ_{ik}$ is the sum of $i$-th row in $Q$. This sampling strategy has been shown to be effective in improving recommendation efficiency and performance in UltraGCN. This strategy removes edges in the item-item graph, which actually follows a similar paradigm to the process of finding graph cut sets. Therefore, to explore the characteristics of the sampled item-item graph in UltraGCN, we measure the algebraic connectivity of the sampled graphs that retain different number of neighbors. The algebraic connectivity, which is also known as the Fielder value, is defined as the second-smallest eigenvalue derived by solving the generalized eigenvalue system
\begin{equation}
(I-\tilde{Q})x=\lambda x.
\end{equation}
Here, the term $I-\tilde{Q}$ is also known as the normalized graph Laplacian matrix. Figure \ref{Fig:Fielder} shows the Fielder value of two datasets \textit{Yelp2018} and \textit{Gowalla} on the sampled item-item graphs. As the number of neighbors sampled decreases, the connectivity of the item-item graph gradually decreases, showing a tendency to produce partitions. When the number of sampling neighbors in the \textit{Gowalla} dataset is between 5 and 10 (the interval typically set in UltraGCN), the connectivity of the graph becomes 0. This illustrates that two or more groups of items appear in item-item graph and are not associated with each other. This phenomenon reveals a potential relationship between graph sampling applied to the CF task and graph partitioning techniques.

Next, we show that the graph sampling strategy in UltraGCN is a local approximation of graph partitioning. Finding a cut-set in the graph, which is known as the graph partitioning problem, has been widely searched in the graph theory. An effective approach to the graph partitioning problem is modularity maximization \cite{Newman2004}. The modularity of an undirected weighted graph $\mathcal{G}=\{\mathcal{V}, \mathcal{E}\}$ is defined as
\begin{equation}
M(\mathcal{G})=\frac{1}{2W}\sum \limits^{\mathcal{E}}_{i, j}[A_{ij}-\frac{d_id_j}{2W}]\mathbf{1}_{c_i= c_j},
\label{Eq:Modularity}
\end{equation}
where $W$ is the sum of all edge weights in $\mathcal{G}$, $A_{ij}$ is the edge weight between node $i$ and node $j$, $d_i$ and $d_j$ are degrees of $i$ and $j$, respectively. $\mathbf{1}_{c_i= c_j}$ is the indicator function which equals 1 when $i$ and $j$ belong to the same partition and equals 0 otherwise. 

Consider that there is a subgraph $\mathcal{G}_i$ that contains node $i$ and all its neighbors in the item-item adjacency graph. Edges exist only between node $i$ and its neighboring nodes, where edge weights equal to $\omega_{ij}$ defined in Eq.  \eqref{Eq:GP_UltraGCN_Omega}. In $\mathcal{G}_i$, the total edge weights $W$ equals to the degree of node $i$, and the degree of the node other than node $i$ is equal to the edge weight between it and node $i$. Then Eq. \eqref{Eq:Modularity} can be simplified to
\begin{equation}
M(\mathcal{G}_i)=\frac{1}{2d_i}\sum \limits_{i, j}[\omega_{ij}-\frac{d_j}{2}]\mathbf{1}_{c_i= c_j}=\frac{1}{4d_i}\sum \limits_{i, j}\omega_{ij}\mathbf{1}_{c_i= c_j}.
\end{equation}
Then, the graph filtering strategy Eq. \eqref{Eq:GP_UltraGCN_Filtering} is equivalent to iterative modularity maximization graph partitioning in the above subgraph. In each iteration, the edge with the smallest weight will be removed until node $i$ has $k$ neighbors left.

The above derivation shows that graph sampling in UltraGCN is an approximation of graph partitioning. It shows that items in the recommendation task may have a group structure, which can be exploited to improve the recommendation accuracy. For this phenomenon, it is also easy to find examples in real-world scenarios, such as the cuisine of restaurants and the location of hotels. This provides the idea of achieving accurate recommendation by fine-grained modeling of the intra-group relationships in the partitioned item-item graph. 

On the other hand, the results of Figure \ref{Fig:Fielder} show that the graph sampling strategy in UltraGCN does not always produce partitions. Also, this strategy samples a fixed number of neighbors per item, which may cause selection bias because items have varied adjacent characteristics in real-world scenarios. These limitations motivate us to explore approaches to achieve effective recommendations with partitioned item-item graph, and further propose FPSR model. 

\section{Methodology}
\begin{figure}[t!]
  \centering
  \includegraphics[width=3.1in]{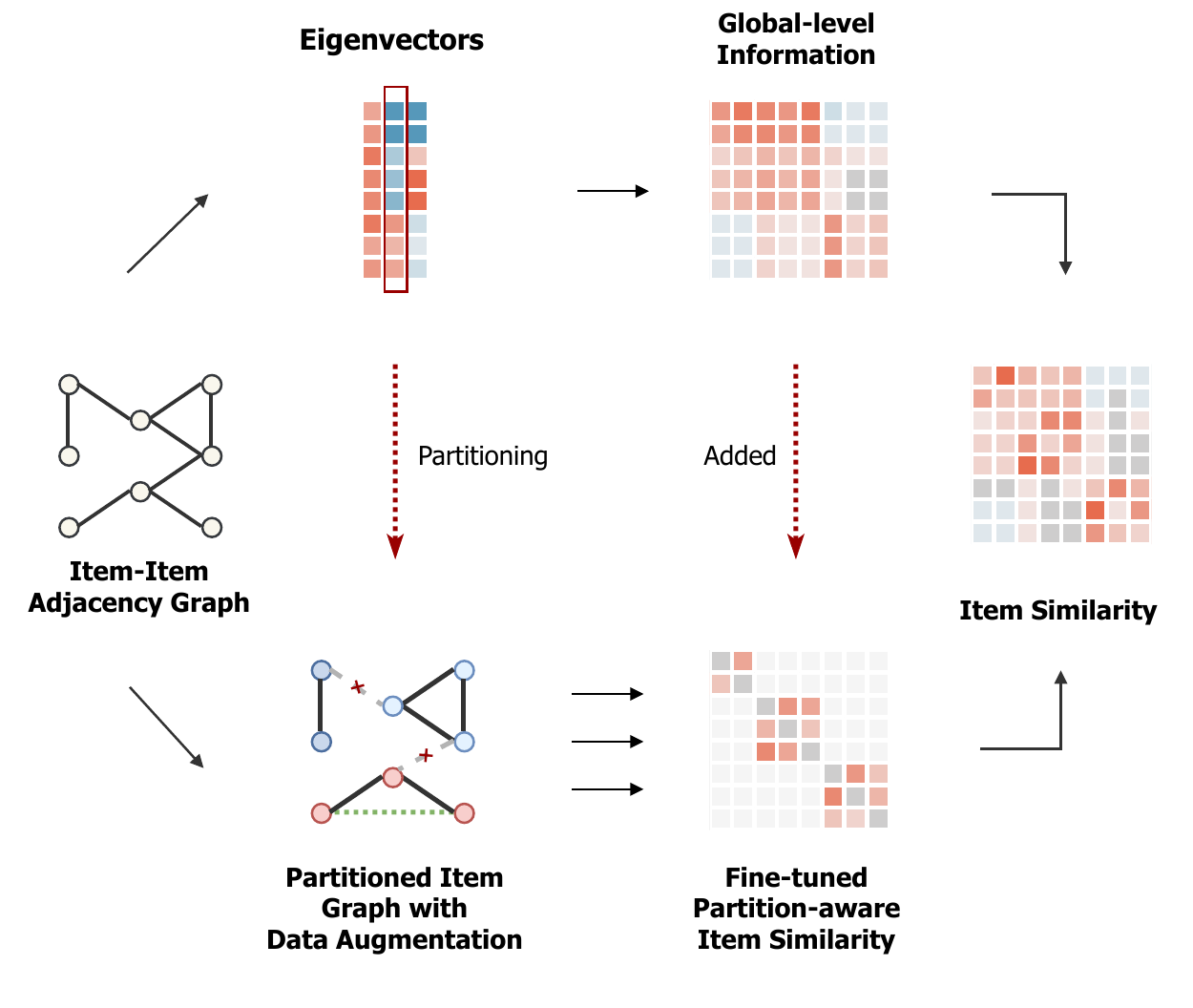}
  \caption{Illustration of the proposed FPSR model. Eigenvectors decomposed from item-item graph are adopted to partition the graph and form global-level information. Item similarities are fine-tuned within each partition to model item relationships together with global information.}
  \Description{Six components: Item-item Adjacency Graph, Eigenvectors, Global-level Information, Partitioned Item Graph with Data Augmentation, Fine-tuned Partition-aware Item Similarity, and Item Similarity. From Item-item Adjacency Graph, Eigenvectors is generated and act together with Item-item Adjacency Graph to generate Partitioned Item Graph with Data Augmentation. Eigenvectors is then used to generate Global-level Information, which is acted on producing Fine-tuned Partition-aware Item Similarity together with the Partitioned Item Graph with Data Augmentation. Finally, both Global-level Information and Partition-aware Item Similarity are used to produce Item Similarity for the final recommendation.}
  \label{Fig:Illu}
  \end{figure}

In this section, we propose our FPSR model as illustrated in Figure \ref{Fig:Illu}. We also perform a complexity analysis to show the advantages of the proposed FPSR in terms of efficiency and scalability compared to existing methods.

\subsection{Recursive Spectral Graph Partitioning}
First, we will divide the item set $\mathcal{I}$ into multiple groups through graph partitioning. In Section \ref{Subsec:GraphCF}, we discuss the Fielder value of the item-item graph, which is defined as an eigenvalue of the normalized Laplacian matrix. In graph theory, the corresponding eigenvector of the Fielder value has been proven to be a good approximation of modularity maximization graph partitioning \cite{Newman2013}. Although this eigenvector can be derived through the eigendecomposition of $I-\tilde{Q}$, the high density of $\tilde{Q}$ can easily lead to storage and efficiency problems. Instead, we can perform truncated singular value decomposition (SVD) on the highly sparse matrix $\tilde{R}$ to find its right singular vector with the second-largest singular value, because $\tilde{Q}$ is constructed by $\tilde{R}^T\tilde{R}$. Several efficient algorithms are designed to solve this problem, like Lanczos method \cite{Baglama2005} and LOBPCG \cite{Knyazev2001}. Suppose $\mathbf{v}$ the right singular vector with second-largest singular value of $\tilde{R}$, which is also known as Fielder vector of $I-\tilde{Q}$, the partition of node $i$ can be derived through
\begin{equation}
i\in
\begin{cases}
    \mathcal{I}_1, & sign(\mathbf{v}_{i})=+1, \\
    \mathcal{I}_2, & sign(\mathbf{v}_{i})=-1.
\end{cases}
\label{Eq:Partitioning}
\end{equation}
The above strategy divides the item set into two partitions without specifying the size of the partition. As the partition size will significantly affect the computational complexity of intra-partition item similarity modeling, it is usually expected to be less than a specified limit. Therefore, we introduce a size ratio $\tau$. When the divided item set $\mathcal{I}_n$ satisfies $\lvert \mathcal{I}_n \rvert/\lvert \mathcal{I}\rvert \geq \tau$, we recursively apply the graph partitioning strategy Eq. \eqref{Eq:Partitioning} to $\mathcal{I}_n$. This ensures all divided item sets have limited size proportional to the entire item set. The selection of $\tau$ will be discussed through experimental analysis in Section \ref{Subsec:Tau}.

\subsection{\textls[-7]{Fine-tuning Intra-partition Item Similarities}}
Through recursively performing partitioning on the item-item graph, we can derive several item groups.  Items within the same cluster are usually considered more closely related, so item similarity modeling within groups should be given more attention. On the other hand, the similarities between items in different partitions should not be ignored. Since the results of graph partitioning are not always ideal, the recommendation model should retain the ability to model item similarities across partitions. Therefore, it becomes a key challenge to optimize the item similarity modeling process by combining global and local relationships inside and outside the partitions. 

Suppose $C\in\mathbb{R}^{\lvert \mathcal{I} \rvert \times \lvert \mathcal{I} \rvert}$ is a item similarity matrix for the entire item set, a conventional learning process of $C$ is conducted  by encoding $R$ as \cite{Steck2019}: 
\begin{equation}
\mathop{\arg\min}\limits_{C}\  \lVert R - R C\rVert^2 + \frac{\theta_2}{2}\lVert C \rVert_F^2,\ \ s.t. diag(C)=0,
\label{Eq:Encoding}
\end{equation}
where the diagonal zero constraint is introduced to prevent trivial solutions. Although it is feasible to simply apply this encoding problem within each partition of the item-item graph, the problems discussed above remains unresolved. Next, we show how our proposed FPSR model achieves intra-partition similarity modeling by combining global-level information in item-item graph and the prior knowledge of partitions.

\subsubsection{Global-level Information}
The Fielder vector derived by performing truncated SVD on $\tilde{R}$ has acted on the item-item graph partitioning. In addition to the sign of the elements in the vector, the numerical values can also represent the similarity relationships between the corresponding nodes, which has been shown in existing research \cite{Shi2000}. This means the global item similarity can be approximated efficiently by the outer product of the Fielder vector. Suppose the corresponding values in Fielder vector of item $i$ and four other items are $\mathbf{v}_i, \mathbf{v}_1, \mathbf{v}_2, \mathbf{v}_3$, and $\mathbf{v}_4$, which satisfy $\mathbf{v}_1<\mathbf{v}_i<\mathbf{v}_2<0<\mathbf{v}_3<\mathbf{v}_4$ and $\mathbf{v}_2 - \mathbf{v}_i > \mathbf{v}_i - \mathbf{v}_1$. Then, the outer products of the Fielder vector have the following property:
\begin{equation}
\mathbf{v}_1\mathbf{v}_i>\mathbf{v}_2\mathbf{v}_i>0>\mathbf{v}_3\mathbf{v}_i>\mathbf{v}_4\mathbf{v}_i.
\end{equation}
It is shown that the similarities across different partitions are correctly ordered, but not always within the same partition. Item 2 has a Fielder vector value closer to item $i$ than item 1, but it has a smaller outer product. It suggests that the Fielder vector can capture global item similarity features, especially for items between different partitions. This fits in with our objective, which is to fine-tune the item similarity within the partition. In addition to the Fielder vector, existing studies \cite{Shi2000, Newman2013} also suggest that other top eigenvectors also contain the information for partitioning, which is demonstrated by experimental analysis. Therefore, we include the top-$k$ right singular vectors of $\tilde{R}$ in modeling item similarities, denoted as $V\in \mathbb{R}^{\mathbf{\lvert \mathcal{I} \rvert} \times k}$.  As shown in \cite{Chen2021}, $V$ is the solution of the following low-rank factorization problem:

\begin{equation}
\mathop{\arg\min}\limits_{U, V} \lVert \tilde{R}-UV^T \rVert_F^2,\ s.t. V^TV=I.
\end{equation}
And the approximation of $R$ can be derived by
\begin{equation}
D_U^{\frac{1}{2}}UV^TD_I^{\frac{1}{2}}=D_U^{\frac{1}{2}}U(V^TV)V^TD_I^{\frac{1}{2}}=R D_I^{-\frac{1}{2}}VV^TD_I^{\frac{1}{2}}.
\end{equation}
In FPSR, we set $W=D_I^{-\frac{1}{2}}VV^TD_I^{\frac{1}{2}}$, and construct the item similarity matrix as
\begin{equation}
C=\lambda W+S,
\label{Eq:C}
\end{equation}
where $S$ is the intra-partition similarity matrix to be fine-tuned. 

\subsubsection{Local Prior Knowledge of Partitions}
When the graph is partitioned, the item similarity will be fine-tuned in each partition. Here, partitions can be considered as prior knowledge that excludes some less relevant items and reduces noise in similarity modeling. However, this prior knowledge does not act explicitly on the fine-tuning of intra-partition similarities, which is still based on the encoding of the interaction matrix. Here, we propose a strategy to incorporate partition information as prior knowledge. Suppose that $\mathbf{1}_n \in \mathbb{R}^{1\times \lvert \mathcal{I} \rvert}$ is an interaction vector defined as
\begin{equation}
\mathbf{1}_{n, i}=
\begin{cases}
    1, & \text{if }i \in \mathcal{I}_n,\\
    0, & \text{otherwise.}
\end{cases}
\end{equation}
Such interaction vector can be considered as a data augmentation of the original interaction matrix $R$, which is equivalent to a user who interacts with all items within a partition.

On the basis of Eq. \eqref{Eq:C} and the proposed local partition-aware interaction vector, we finally formulate the intra-partition similarity modeling problem as
\begin{equation}
\begin{aligned}
\underset{S}{\mathrm{argmin}}&\ \frac{1}{2}\lVert R - R(\lambda W+S)\rVert_F^2+\frac{\theta_2}{2}\lVert D_I^{\frac{1}{2}}(\lambda W+S) \rVert_F^2 \\
&+\theta_1 \lVert S \rVert_1 + \sum \limits_{n} \frac{\eta}{2} \lVert \mathbf{1}_n^T-\mathbf{1}_n^T(\lambda W+S) \rVert_F^2\\
s.t.&\ diag(S)=0,\ S\geq0,\ S_{ij|\mathcal{G}(i) \neq \mathcal{G}(j)}=0,
\end{aligned}
\label{Eq:Problem}
\end{equation}
where $\mathcal{G}(i)$ is the partition item $i$ assigned. $\lVert D_I^{\frac{1}{2}}(\lambda W+S) \rVert_F^2$ is the $l_2$ regularization term weighted by column based on the occurrence in $R$. The $l_1$ regularization term and non-negative constraint are added to learn a sparse solution with the aim of reducing the number of parameters to achieve efficient recommendations. When $C$ is obtained, the top-$K$ recommendations for user $u$ is generated as
\begin{equation}
    \overset{K}{\mathrm{argmax}}\ \mathbf{r}_u C,
    \label{Eq:Recommend}
\end{equation}
where $\mathbf{r}_u$ is the $u$-th row of $R$.

\subsubsection{Optimization}
With the constraint $S_{ij|\mathcal{G}(i) \neq \mathcal{G}(j)}=0$, the problem Eq. \eqref{Eq:Problem} can be divided to several sub-problem in each partition. Here we denote the item-item similarity matrix in partition $n$ as $S_n$, and $S$ is derived by concatenation of similarity matrix in all partitions:
\begin{equation}
S=diag(S_1, ..., S_n).
\label{Eq:FullS}
\end{equation}
The constraint optimization problem in a partition can be solved using the alternate direction of multiplier method (ADMM) \cite{Boyd2011, Steck2020WSDM} by converting the problem Eq. \eqref{Eq:Problem} to
\begin{equation}
\begin{aligned}
\underset{Z_n, S_n}{\mathrm{argmin}}\ &\frac{1}{2}tr(Z_n^T\hat{Q}Z_n)-tr((I-\lambda W_n)\hat{Q}Z_n)
+\theta_1 \lVert S_n \rVert_1\\
&s.t.\ diag(Z_n)=0,\ S_n\geq0,\ Z_n=S_n,
\end{aligned}
\label{Eq:Problem_ADMM}
\end{equation}
where $tr(\cdot)$ denotes the trace of matrix, and $\hat{Q}$ is defined as
\begin{equation}
\hat{Q}=R_n^TR_n+\theta_2D_I+\eta.
\label{Eq:Qhat}
\end{equation}
where $W_n$ and $R_n$ are obtained by selecting only the items of partition $n$ in $W$ and $R$ respectively. Then, the updating rules of the $t+1$-th iteration are derived as
\begin{equation}
\begin{aligned}
Z_n^{(t+1)}=&(\hat{Q}+\rho I)^{-1}(\hat{Q}(I-\lambda W_n)\\
&+\rho (S_n^{(k)}-\Phi_n^{(k)})-diagMat(\mathbf{\mu})),\\
S_n^{(t+1)}=&(Z_n^{(t+1)}+\Phi_n^{(k)}-\frac{\theta_1}{\rho})_+,\\
\Phi_n^{(t+1)}=&\Phi_n^{(k)}+Z_n^{(t+1)}-S_n^{(t+1)}.
\end{aligned}
\label{Eq:ADMM}
\end{equation}
Here, $\Phi_n$ is the dual variable,  $\rho$ is the hyperparameter introduced by ADMM, and $\mathbf{\mu}$ is the vector of the augmented Lagrange multiplier defined as
\begin{equation}
\begin{aligned}
\mu=&\ diag((\hat{Q}+\rho I)^{-1}(\hat{Q}(I-\lambda W_n)\\
&+\rho (S_n^{(k)}-\Phi_n^{(k)}))) \oslash diag((\hat{Q}+\rho I)^{-1}),
\end{aligned}
\end{equation}
where $\oslash$ denotes the element-wise division. When optimization is finished, $S_n$ is returned as the item-item weight matrix for this partition. In the actual optimization process, we filter out small values in the sparse matrix $S_n$ after the optimization to reduce the noise. Generally, setting the filtering threshold between \textit{1e-3} and \textit{5e-3} is found to be effective in filtering out noisy small values and has no statistically significant effect on the performance of the recommendation. 

\subsection{Computational Complexity}
Next, we conduct theoretical analysis to illustrate the major highlight of FPSR: the efficiency and scalability improvement in both time and storage brought by the graph partitioning.

The major contributions to computational complexity in FPSR are the generation of global-level similarity matrix $W$ and the optimization of partitioned intra-partition similarity matrix $S$. $W$ is generated by the top-$k$ eigenvectors on $\tilde{Q}$ through truncated SVD performed on the sparse matrix $\tilde{R}$, yielding the computational complexity of $\mathcal{O}(T(k^3+k\lvert \mathcal{T} \rvert))$ \cite{Shen2021, Journee2010}, where $T$ denotes the iteration numbers and $\lvert \mathcal{T} \rvert \gg k^2$ is the nonzero element numbers in $R$. Because only the matrix $\tilde{R}$ is required in truncated SVD instead of $\tilde{Q}$, the storage cost is $\mathcal{O}(\lvert \mathcal{T} \rvert)$. On the other hand, FPSR requires computing the first 2 eigenvectors of the item adjacency matrix at each time of partitioning.  As only the first 2 eigenvalues are required, which is usually much smaller than $k$, its contribution to the overall computational and storage cost can be neglected. 

Optimization of global item similarity like Eq. \eqref{Eq:Encoding} requires a computational complexity of at least $\mathcal{O}(\lvert \mathcal{I} \rvert^{2.376})$ contributed by the matrix inverse \cite{Steck2019, Steck2020WSDM}, and a storage cost of $\mathcal{O}(\lvert \mathcal{I} \rvert^2)$. In FPSR, the optimization is performed within each partition, whose size can be explicitly limited with the hyperparameter $\tau$. As the maximum size of each partition is $\tau \lvert \mathcal{I} \lvert$, the upper bound of the total computational complexity is $\mathcal{O}(\lvert \mathcal{I}\rvert^{2.376} \tau^{1.376})$, and the storage cost is $\mathcal{O}(\lvert \mathcal{I} \rvert^2 \tau)$. These are rough upper bounds that assume all partitions to be the same size $\tau\lvert \mathcal{I} \rvert$. The actual partitions produced usually have uneven sizes smaller than $\tau\lvert \mathcal{I} \rvert$, yielding lower computing and storage costs. 

\section{Experiments}
\subsection{Experimental Setup}
\subsubsection{Datasets and Evaluation Metrics}
We carry out experiments on four public datasets: \textit{Amazon-cds}, \textit{Douban}, \textit{Gowalla}, and \textit{Yelp2018}. The same data split as existing studies on CF models \cite{Wang2019, He2020, Mao2021} is adopted to ensure a fair comparison. The statistics of all 4 datasets are summarized in Table \ref{Tab:Dataset}. For evaluation metrics, two metrics that are popular in the evaluation of CF tasks are involved: NDCG@$K$ and Recall@$K$, where $K$ is set to 20. 

\begin{table}[t!]
\renewcommand{\arraystretch}{0.75}
\caption{Statistics of datasets}
\label{Tab:Dataset}
\begin{center}
\begin{tabular}{@{}c|cccc@{}}
\toprule
Dataset & \multicolumn{1}{c}{\#User} & \multicolumn{1}{c}{\#Item} & \multicolumn{1}{c}{\#Interaction} & \multicolumn{1}{c}{Density} \\ \midrule
Amazon-cds & 43,169 & 35,648 & 777,426 & 0.051\% \\
Douban & 13,024 & 22,347 & 792,062 & 0.272\% \\
Yelp2018 & 31,668 & 38,048 & 1,561,406 & 0.130\% \\
Gowalla & 29,858 & 40,981 & 1,027,370 & 0.084\% \\ \bottomrule
\end{tabular}
\end{center}
\end{table}

\subsubsection{Baselines}
We compare our proposed model with several types of CF models:
\begin{itemize}
    \item Non-parametric models: Item-KNN \cite{Deshpande2004}, GF-CF \cite{Shen2021}.
    \item MF model: MF-BPR \cite{Rendle2009}.
    \item GCN models: LightGCN \cite{He2020}, UltraGCN \cite{Mao2021}, SimGCL \cite{Yu2022}.
    \item Item similarity models: SLIM \cite{Ning2011}, EASE \cite{Steck2019}, BISM \cite{Chen2020}.
\end{itemize}
All models are implemented and tested with RecBole \cite{Zhao2021} toolbox to ensure the fairness of performance comparison.

\begin{table*}[t!]
\setlength{\tabcolsep}{3.75pt}
\renewcommand{\arraystretch}{0.75}
\setul{1pt}{.4pt}
\caption{Performance comparison on 4 public datasets}
\label{Tab:PerformancePublic}
\begin{center}
\begin{tabular}{@{}cccccccccccc@{}}
\toprule
\multicolumn{2}{c}{Model} & ItemKNN & GF-CF & BPR & LightGCN & SimGCL & UltraGCN & EASE & SLIM & BISM & FPSR \\ \midrule
\multirow{4}{*}{Amazon-cds} & Recall@20 & 0.1331 & 0.1350 & 0.1111 & 0.1346 & 0.1468 & 0.1487 & 0.1433 & 0.1460 & \ul{0.1541} & \textbf{0.1576*} \\
 & NDCG@20 & 0.0756 & 0.0725 & 0.0569 & 0.0701 & 0.0781 & 0.0806 & 0.0845 & 0.0807 & \ul{0.0874} & \textbf{0.0896*} \\ \cmidrule(l){2-12} 
 & \#Parameters & - & - & 5.04M & 5.04M & 5.04M & 5.04M & 1271M & 73.9M & 111M & 3.87M \\
 & Training Time (s) & 24 & 26 & 268 & $3.9\times 10^4$ & 526 & $3.3\times 10^3$ & 37 & 561 & 397 & 50 \\ \midrule
\multirow{4}{*}{Yelp2018} & Recall@20 & 0.0563 & \ul{0.0697} & 0.0576 & 0.0653 & 0.0681 & 0.0683 & 0.0657 & 0.0644 & 0.0662 & \textbf{0.0703*} \\
 & NDCG@20 & 0.0469 & \ul{0.0571} & 0.0468 & 0.0532 & 0.0556 & 0.0561 & 0.0552 & 0.0542 & 0.0559 & \textbf{0.0584*} \\ \cmidrule(l){2-12} 
 & \#Parameters & - & - & 4.46M & 4.46M & 4.46M & 4.46M & 1448M & 126M & 191M & 3.27M \\
 & Training Time (s) & 39 & 23 & 208 & $7.4\times 10^4$ & $2.9\times 10^3$ & 617 & 22 & $1.1\times 10^3$ & 783 & 35 \\ \midrule
\multirow{4}{*}{Douban} & Recall@20 & 0.1923 & 0.1719 & 0.1347 & 0.1571 & 0.1699 & 0.1925 & 0.2038 & 0.2002 & \textbf{0.2158} & \ul{0.2095} \\
 & NDCG@20 & 0.1686 & 0.1365 & 0.0966 & 0.1206 & 0.1346 & 0.1556 & 0.1786 & 0.1731 & \ul{0.1889} & \textbf{0.1950*} \\ \cmidrule(l){2-12} 
 & \#Parameters & - & - & 5.04M & 5.04M & 5.04M & 5.04M & 499M & 103M & 158M & 2.14M \\
 & Training Time (s) & 14 & 19 & 441 & $2.2\times 10^4$ & $2.7\times 10^3$ & $5.9\times 10^3$ & 16 & 226 & 410 & 33 \\ \midrule
\multirow{4}{*}{Gowalla} & Recall@20 & 0.1246 & 0.1849 & 0.1627 & 0.1820 & 0.1762 & \ul{0.1862} & 0.1765 & 0.1699 & 0.1724 & \textbf{0.1884*} \\
 & NDCG@20 & 0.0907 & 0.1518 & 0.1378 & 0.1547 & 0.1495 & \textbf{0.1580} & 0.1467 & 0.1382 & 0.1443 & \ul{0.1566} \\ \cmidrule(l){2-12} 
 & \#Parameters & - & - & 4.53M & 4.53M & 4.53M & 4.53M & 1679M & 84.6M & 127M & 3.44M \\
 & Training Time (s) & 34 & 21 & 147 & $5.2\times 10^4$ & $4.4\times 10^3$ & $3.8\times 10^3$ & 20 & $1.7\times 10^3$ & 683 & 87 \\ \bottomrule
\end{tabular}
\end{center}
In each metric, the best result is \textbf{bolded} and the runner-up is \ul{underlined}. * indicates the statistical significance of $p<0.01$.
\end{table*}

\subsubsection{Parameter Settings}
The hyperparameters of FPSR and compared baseline models are explored within their hyperparameter space. A five-fold cross-validation is performed to select the hyperparameters with the best performance in the validation set. For FPSR, we set the number of eigenvectors extracted to 256 to maintain consistency with GF-CF. The $l_1$ and $l_2$ regularization weights $\theta_1$ and $\theta_2$ are tuned in [0.1, 0.2, 0.5, 1, 2, 5], $\lambda$ and $\tau$ are tuned in [0.1, 0.2, 0.3, 0.4, 0.5], and $\eta $ is tuned in [0.01, 0.1, 1]. For other baselines, we carefully tune the hyperparameters to achieve the best performance. To keep consistency, for MF-BPR and GCN-based models, the embedding size for users and items is set to 64, the learning rate is set to $10^{-3}$, and the training batch size is set to 2048. For SLIM, BISM, and FPSR, the sparsity of the learned similarity matrix cannot be set explicitly, but can be changed by adjusting the hyperparameter that controls $l_1$ regularization. Therefore, we list the number of model parameters with the performance to make a comprehensive comparison. All sparse matrices are stored in compressed sparse row (CSR) format, which contains approximately twice the parameter numbers as the number of non-zero values (NNZ) in the sparse matrix.

\subsection{Performance Comparison}
As the different models vary in computation and storage cost, we report the number of parameters and training time together with the evaluation metrics. All experiments are conducted on with the same Intel(R) Core(TM) i9-10900X CPU @ 3.70GHz machine with one Nvidia RTX A6000 GPU. Table \ref{Tab:PerformancePublic} reports the performance comparison on 4 public datasets. Based on Table \ref{Tab:PerformancePublic}, the highlights are summarized as follows.

First, the proposed FPSR model achieves the overall best performance on 4 datasets. The results of statistical analysis indicate that FPSR achieves a significant improvement in all metrics of \textit{Amazon-cds} and \textit{Yelp2018} datasets, NDCG@20 on \textit{Douban} dataset, and Recall@20 on \textit{Gowalla} dataset. For Recall@20 on \textit{Douban} dataset, FPSR achieves runner-up result with less than 10\% of the parameter storage and training time compared to BISM. For NDCG@20 on \textit{Gowalla} dataset, FPSR achieves competitive result compared to UltraGCN with 50 times faster training speed and fewer parameters.

Second, FPSR shows a strong ability to adapt to different datasets. The results in Table \ref{Tab:PerformancePublic} show that the baselines perform differently in various datasets. GF-CF and GCN models (LightGCN, SimGCL and UltraGCN) significantly outperform EASE, SLIM and BISM in \textit{Gowalla} dataset, while opposite trends are shown in \textit{Amazon-cds} and \textit{Douban} datasets. In a recent study \cite{Chin2022}, the authors discuss the selection bias of the datasets on recommendation task and cluster the publicly used datasets into different groups. Our experiments verify that this difference exists between different groups of datasets (e.g. \textit{Amazon-cds} and \textit{Gowalla}). Meanwhile, FPSR performs well on various groups of datasets, demonstrating the strong ability of FPSR in fitting datasets with different characteristics.

Finally, the advantage of FPSR appears more significant when considering with the number of parameters and training time. Existing GCN models are iteratively optimized by small-batch gradient descent, resulting in relatively longer training times. In contrast, learning item similarities on the entire dataset is faster, but also yields tens or even hundreds of times the growth of the training parameters. In FPSR, benefiting from the identification and discovery of partition structures in item-item graph, similarity can be fine-tuned in the local partition while retaining global-level information. As a result, the size of the similarity modeling problem can be controlled by $\tau$, which leads to substantial parameter storage savings compared to item similarity models, and high training speed compared to GCN models.

\subsection{Ablation Analysis}
To validate the contribution of each component in the proposed FPSR model, we conduct experiments on 3 variants of FPSR listed as follows:

\begin{itemize}
    \item FPSR ($\eta=0$): it removes the local prior knowledge term from the optimization problem of $S$.
    \item FPSR ($\lambda=0$): it removes the global information $W$ and makes $C=S$, which only contains item similarities in each partition.
    \item FPSR ($\theta_2=0$): it removes the $l_2$ regularization term from the optimization problem of $S$.
\end{itemize}

Here we do not set $\theta_1$ to 0 as a variant, because it is necessary to guarantee the sparsity of the learned local item-item similarity matrix in FPSR. Table \ref{Tab:Ablation} reports the results of the ablation analysis. It can be seen that FPSR consistently outperforms all variants on the test datasets. On the other hand, comparisons between variants show that different components of FPSR contribute differently on each dataset, revealing distinct sensitivity of the datasets to global and local information.  FPSR($\lambda=0$) shows a large performance degradation in \textit{Gowalla} and \textit{Yelp2018} datasets, while FPSR($\eta=0$) degrades most significantly in \textit{Douban} dataset. In \textit{Amazon-cds} dataset, all three variants show a statistically significant performance degradation compared to the full version of FPSR. This verifies the effectiveness of the individual components in FPSR.

\begin{table}[t!]
\setlength{\tabcolsep}{3.5pt}
\renewcommand{\arraystretch}{0.75}
\setul{1pt}{.4pt}
\caption{Ablation analysis}
\label{Tab:Ablation}
\begin{center}
\begin{tabular}{@{}ccccc@{}}
\toprule
Dataset & \multicolumn{2}{c}{Amazon-cds} & \multicolumn{2}{c}{Douban} \\ \midrule
Metrics & Recall@20 & NDCG@20 & Recall@20 & NDCG@20 \\ \midrule
FPSR ($\eta = 0$) & 0.1540 & 0.0873 & 0.2046 & 0.1909 \\
FPSR ($\lambda = 0$) & \ul{0.1542} & \ul{0.0887} & \ul{0.2085} & \ul{0.1945} \\
FPSR ($\theta_2 = 0$) & 0.1539 & 0.0870 & 0.2084 & 0.1943 \\
FPSR & \textbf{0.1576} & \textbf{0.0896} & \textbf{0.2095} & \textbf{0.1950} \\ \midrule
Dataset & \multicolumn{2}{c}{Gowalla} & \multicolumn{2}{c}{Yelp2018} \\ \midrule
Metrics & Recall@20 & NDCG@20 & Recall@20 & NDCG@20 \\ \midrule
FPSR ($\eta = 0$) & \ul{0.1883} & \textbf{0.1566} & \ul{0.0702} & \ul{0.0582} \\
FPSR ($\lambda = 0$) & 0.1785 & 0.1486 & 0.0662 & 0.0556 \\
FPSR ($\theta_2 = 0$) & 0.1832 & \ul{0.1501} & 0.0692 & 0.0574 \\
FPSR & \textbf{0.1884} & \textbf{0.1566} & \textbf{0.0703} & \textbf{0.0584} \\ \bottomrule
\end{tabular}
\end{center}
\end{table}

\subsection{Detailed Analysis}
\label{Subsec:Param}
\subsubsection{Impact of $\lambda$ on performance}
\begin{figure}[t!]
  \centering
  \includegraphics[width=1.65in]{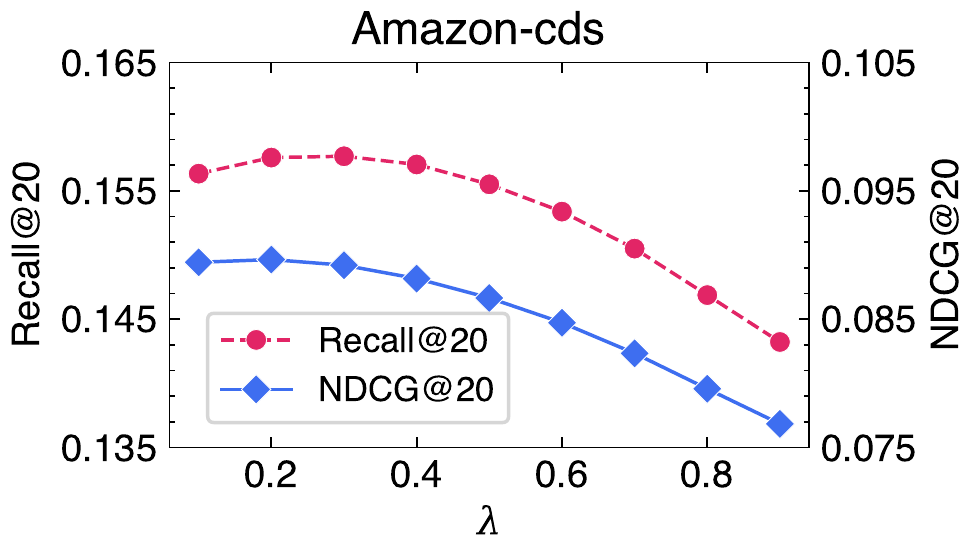}
  \hfil
  \includegraphics[width=1.65in]{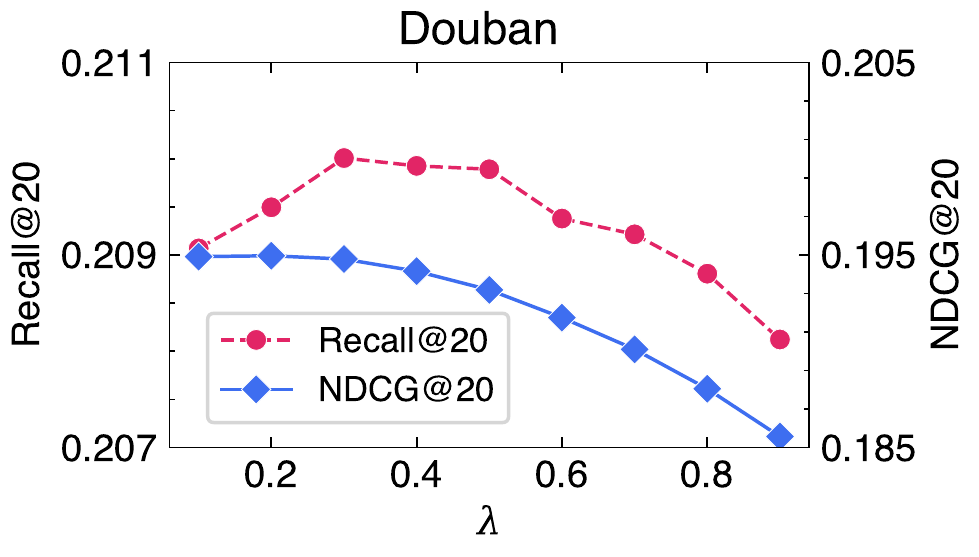}
  \hfil
  \includegraphics[width=1.65in]{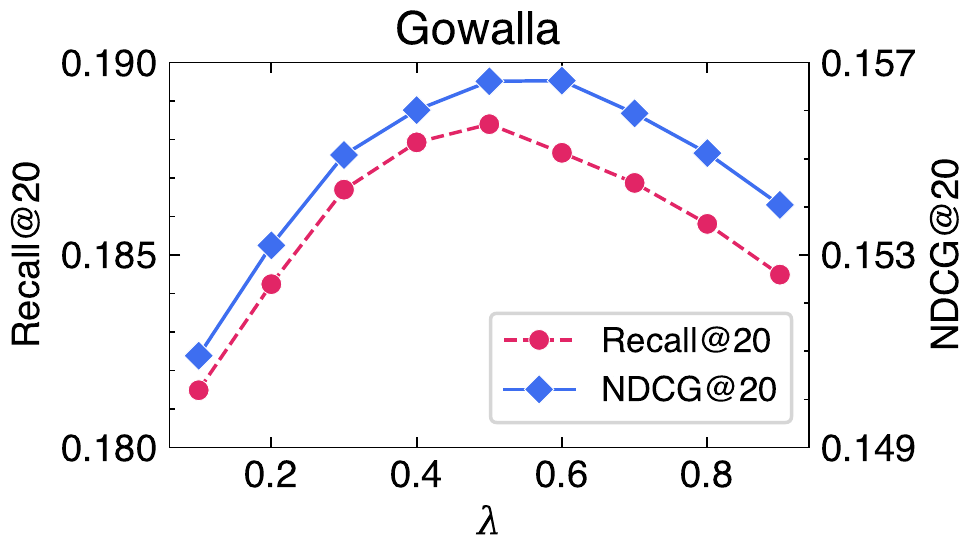}
  \hfil
  \includegraphics[width=1.65in]{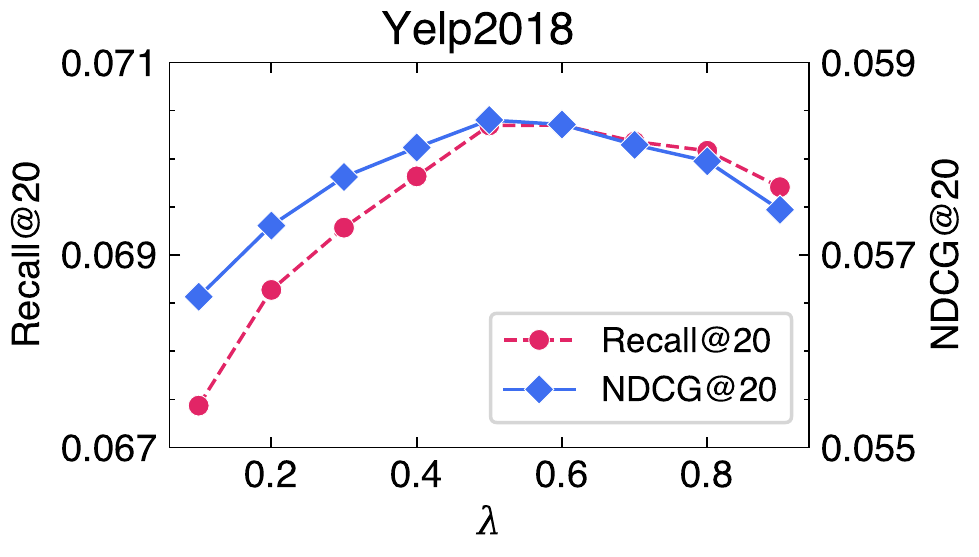}
  \caption{Performance of FPSR with different $\lambda$.}
  \Description{Line graph showing the Recall@20 and NDCG@20 against of the hyperparameter lambda from 0.1 to 0.9 in increments of 0.05 on the X axis. Four subgraphs are shown, corresponding to four test datasets: Amazon-cds, Douban, Gowalla, and Yelp2018. In all subgraphs, both two lines for Recall@20 and NDCG@20 increase first and then decrease when lambda grows from 0.1 to 0.9. Recall@20 and NDCG@20 peaks in different subgraphs is different value of lambda. It is 0.2 in Amazon-cds, 0.3 in Douban, and 0.5 in Gowalla and Yelp2018.}
  \label{Fig:Lambda}
  \end{figure}

To test the impact of $\lambda$ on FPSR, we sett it to different values between 0.1 and 0.9 with a step size of 0.1. Figure \ref{Fig:Lambda} shows the performance in all tested datasets. In summary, as $\lambda$ increases, the performance of FPSR increases and then falls. FPSR maintains the best performance when setting $\lambda$ between 0.2 and 0.6 in all datasets, while different datasets have different sensitivity to small or large values of $\lambda$. This phenomenon confirms the role of $W$ in introducing global-level information to the fine-tuning of item similarities in each partition, while the effectiveness varies depending on the characteristics of the dataset.

\subsubsection{Impact of $\theta_1$ on parameter numbers}
\begin{figure}[t!]
  \centering
  \includegraphics[width=1.71in]{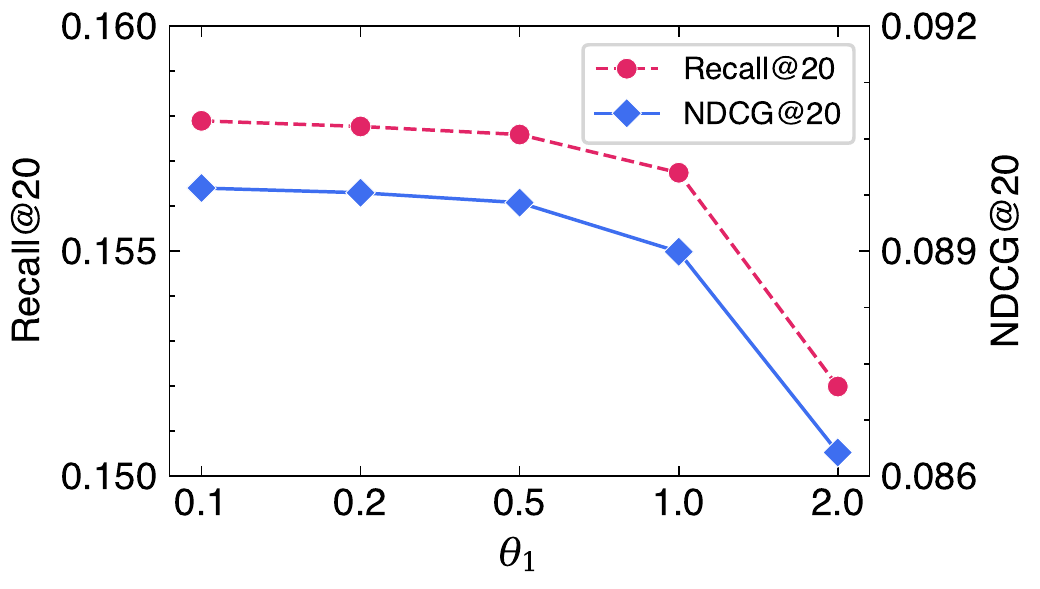}
  \hfil
  \includegraphics[width=1.59in]{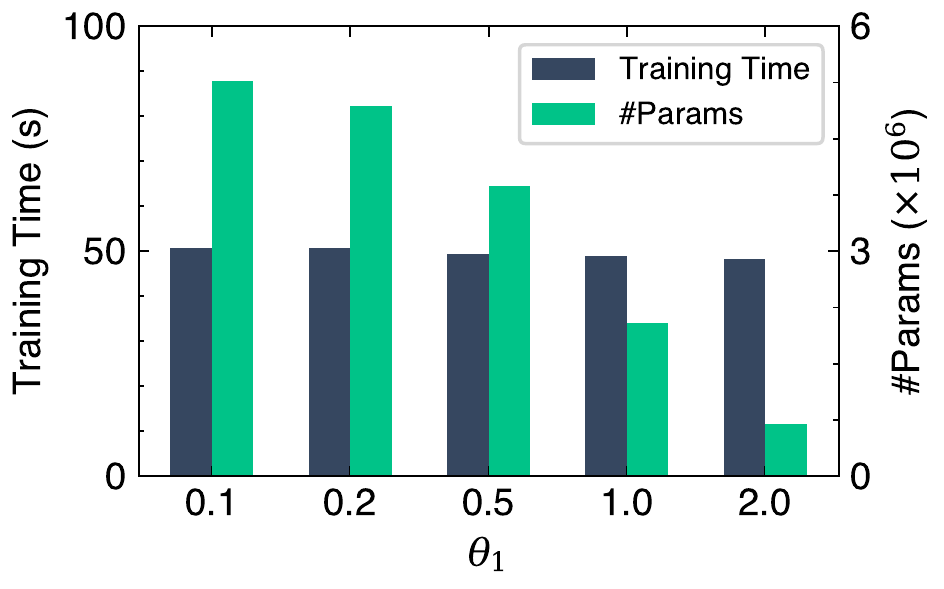}
  \caption{Impact of $\theta_1$ on \textit{Amazon-cds} dataset.}
  \Description{Line and bar graphs showing the Recall@20, NDCG@20, training time and the number of parameters against of the hyperparameter theta 1 from 0.1 to 2.0 on the X axis. Two subgraphs are shown. The left subgraph shows two lines for Recall@20, NDCG@20, which decrease slowly when theta 1 increases from 0.1 to 1.0, and then decrease significantly when theta 1 increases from 1.0 to 2.0. The right subgraph shows two groups of bar for training time and the number of parameters respectively. When theta 1 increases from 0.1 to 2.0, the training time stays at around 50 seconds, and the number of parameters consistently decreases from 5 million to less than 1 million.}
  \label{Fig:Theta1}
  \end{figure}

Here we explore the impact of $\theta_1$ on the number of parameters in FPSR. Figure \ref{Fig:Theta1} shows the performance comparison in \textit{Yelp2018} dataset. As $\theta_1$ gradually decreases from 2.0, the number of parameters increases due to the weakening of $l_1$ regularization constraints. Compared to the significant change in the number of parameters, training time is hardly affected by $\theta_1$. At the same time, the performance of FPSR first rises and then stabilizes. This suggests that a reasonable setting of $\theta_1$ can effectively filter the informative values in item similarity modeling, thus reducing storage costs while maintaining optimal model performance.

\subsubsection{Impact of $\tau$ on training efficiency}
\label{Subsec:Tau}
\begin{figure}[t!]
  \centering
  \includegraphics[width=1.71in]{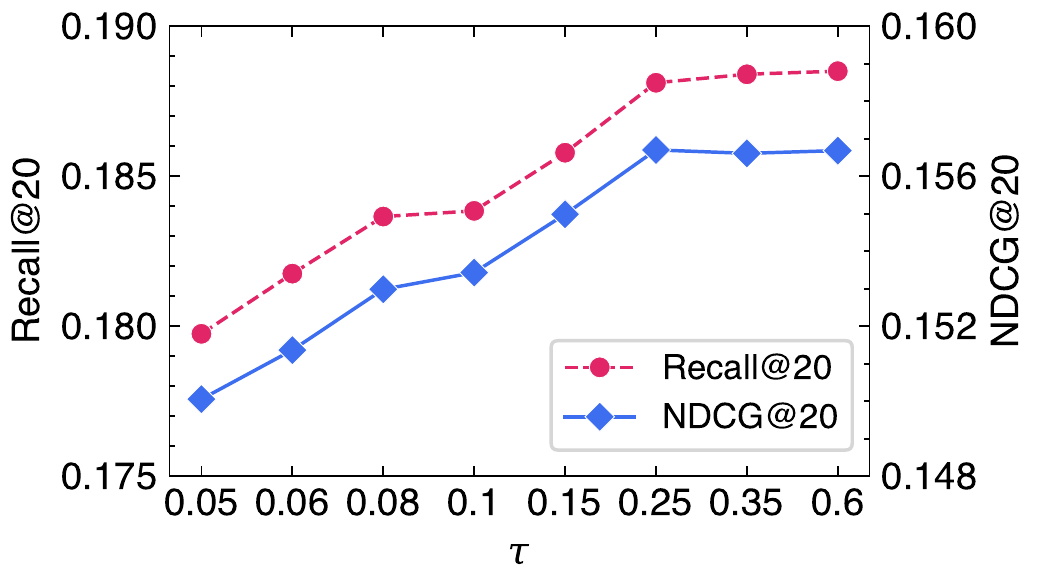}
  \hfil
  \includegraphics[width=1.59in]{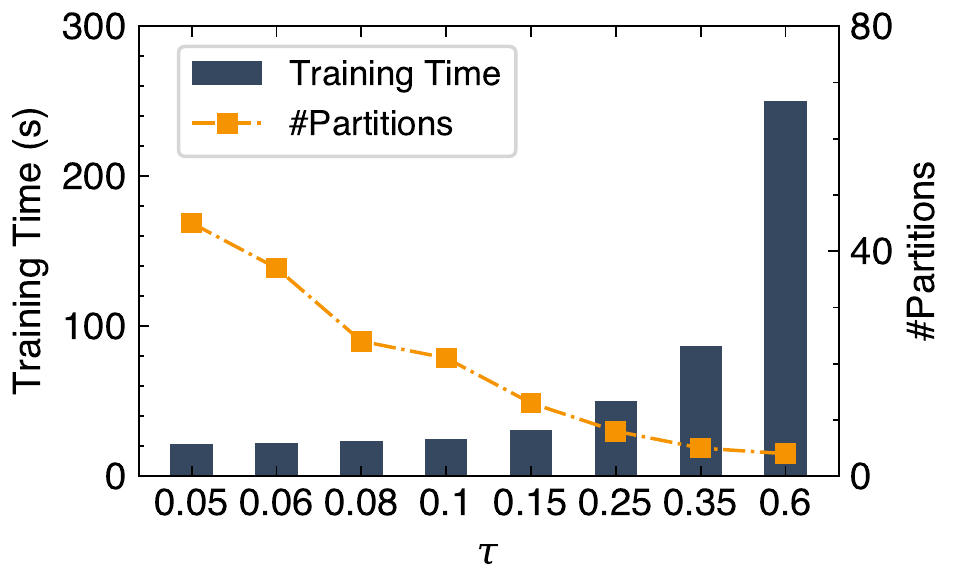}
  \caption{Impact of $\tau$ on \textit{Gowalla} dataset.}
  \Description{Line and bar graphs showing the Recall@20, NDCG@20, training time and the number of partitions against of the hyperparameter tau from 0.05 to 0.6 on the X axis. Two subgraphs are shown. The left subgraph shows two lines for Recall@20, NDCG@20, which increase consistently when tau increases from 0.05 to 0.6. The right subgraph shows a group of bar for training time and a line for the number of partitions respectively. When tau increases from 0.05 to 0.6, the training time increases from 20 seconds to 260 seconds, and the number of partitions decreases from 40 to 5.}
  \label{Fig:SizeLimit}
  \end{figure}

To verify the impact of the recursive partitioning strategy, we test FPSR model by setting $\tau$ to different values. Figure \ref{Fig:SizeLimit} shows the results in \textit{Gowalla} dataset. When $\tau$ is reduced from 0.6 to 0.25, the model performance changes slightly, while the training time is significantly reduced. Continuing to decrease $\tau$, the performance of the model starts to decrease with a rapid increase in the number of partitions. This demonstrates that when $\tau$ is set appropriately, FPSR can significantly improve the training efficiency of the local item similarity fine-tuning process, while maintaining excellent recommendation performance. This observation shows that the graph partitioning strategy in FPSR brings a significant improvement in efficiency, allowing it to be superior in both speed and performance, which are the main concerns in large-scale recommender systems.

\subsection{Case Study}
\begin{figure}[t!]
  \centering
  \includegraphics[width=3.3in]{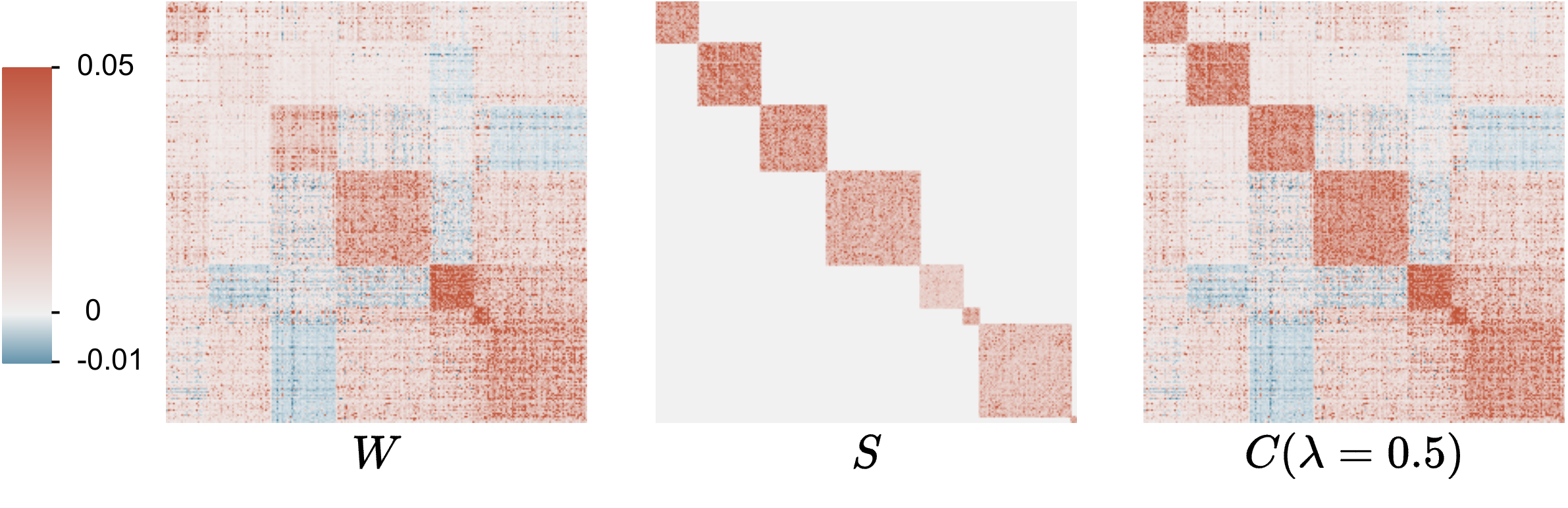}
  \caption{Visualization of $W, S$ and $C$ in \textit{Yelp2018} dataset.}
  \Description{Visualization of matrix W, S, and C, whose values are varied from 0.05 to -0.01. W has positive values for diagonal blocks, and both positive and negative values for other blocks. S only has non-zero values in each block of diagonals, corresponding to the fine-tuned item similarities within each partition. C is the sum of S and lambda times W, where lambda is set to 0.5.}
  \label{Fig:Corr}
  \end{figure}

To further investigate the effect of global-level information in FPSR on the fine-tuning of local item similarity within a partition, we conduct a case study of the similarity matrix $C$ obtained from FPSR. Figure \ref{Fig:Corr} shows visualization of $C$ and its two components $W$ and $S$ learned from \textit{Yelp2018} dataset, containing more than 38k items in total. In the global-level information matrix $W$, the diagonal blocks represent the item similarity within each partition, which correspond to the non-zero regions in $S$. The intra-partition similarities in $W$ are shown to be relatively large compared to the inter-partition similarities, while both of them are varied across partitions. Correspondingly, partitions with relatively high similarity in $W$ show high sparsity in $S$. This demonstrates the benefits of combining global-level information for fine-tuning local item similarity. First, the global-level information preserves item relationships across partitions, providing FPSR with the ability to perform collaborative filtering across the entire item set. Also, it can act on the local item similarity fine-tuning as prior knowledge to facilitate optimization process and reduce the parameter numbers. These factors contribute to great ability of FPSR in recommending items globally and locally within the focused partition, resulting in better performance and scalability compared to existing models.

\section{Related Works}
\subsection{Collaborative Filtering}
Collaborative Filtering (CF) is a widely searched problem in recommender systems. In CF, user-item interactions can be represented as several equivalent forms \cite{Mao2021}: discrete user-item pairs, a sparse user-item interaction matrix, or a user-item bipartite graph. Different treatments of the interactions yield different types of models.

By treating CF task as a matrix completion problem in user-item interaction matrix, matrix factorization (MF) models \cite{Koren2009, Rendle2009} are proposed to learn features of users and items as embedding vectors. Such models enable a lightweight and flexible training process by treating interactions as independent user-item pairs, but neglect the relationships between different entities. In recent years, due to the natural graphical properties of user-item interaction, Graph Convolution Network (GCN) have been widely applied to CF \cite{Wang2019, He2020, Mao2021, Yu2022} to learn user and embeddings through multi-layer graph convolution. In the early proposed NGCF \cite{Wang2019}, heavy designs are used in the standard GCN, including non-linear activation of feature transformation, which has been shown to be unnecessary in CF by LightGCN \cite{He2020}. Subsequent work has been improved based on LightGCN, including the approximation of representation learning \cite{Mao2021} and the introduction of contrastive learning \cite{Lin2022, Yu2022}. 

As an alternative approach to handle user-item interactions, neighborhood-based models \cite{Deshpande2004} treat interactions as the features of users and items to model their similarities.  Ning et al. \cite{Ning2011} first propose sparse linear method (SLIM) to formulate item similarity modeling as an interaction encoding problem. SLIM is then followed by subsequent studies that attempt improve the performance and efficiency, including refining the optimization algorithm \cite{Steck2020WSDM} and applying the denoising module \cite{Steck2020NIPS}.   Compared to MF and GCN models trained small-batch iterative optimization, item similarity models are usually more efficient in modeling global relationships. However, the storage cost of similarity models are highly dependent on the size of the item set, limiting their scalability in practical scenarios with increasing number of items. 

\subsection{CF with Group Structures}
In CF, users are likely to interact with items based on their own preferences and item characteristics, which can lead to the structure of groups or clusters \cite{Heckel2017}. Much of the previous research has investigated the identification of group structures in users and items for the training of embeddings \cite{Heckel2017, Khawar2019}. The derived group structures can also act on the structure learning in neural network \cite{Khawar2020} or the graph sampling in GCN \cite{Liu2021}. Results in such studies reveal the effectiveness of clustering information in enhancing recommendation performance. However, the clustering process in these models has a limited impact on the training of embeddings, whose efficiencies are still constrained by small-batch iterative optimizations. 

Instead, grouping users or items lead to direct change in the scale of the similarity modeling problem, thereby bringing more significant effect on the similarity models. Early works attempt to reduce the computational complexity in learning item similarities by applying clustering methods, but fail to capture global patterns and lead to a trade-off between performance and scalability \cite{OCONNOR1999, Sarwar2002}. Alternatively, LorSLIM \cite{Cheng2014} and BISM \cite{Chen2020} approximate the block diagonal attributes of item similarities in the encoding of the interaction matrix, expecting to detect latent group relationships. Although these efforts demonstrate the benefits of exploiting cluster structure with the superior performance compared to existing models like SLIM, the scalability problem still exists, because the item similarities are still modeled on the entire item set.

Different from existing works, we propose FPSR to focus on the fine-tuning of the item similarities within the partitions of item graph. Both global-level information and the prior knowledge of partitioning are involved in the local similarity modeling, bringing significant improvement in performance, efficiency and scalability compared with state-of-the-art models.

\section{Conclusion and Future Work}
In this paper, we propose a novel efficient and scalable model for recommendation named FPSR, which applies graph partitioning to item-item graphs to fine-tune item similarities within each partitions. FPSR also introduces global-level information to local similarity modeling inside each partition to cope with information loss. Also, we propose a data augmentation strategy to explicitly add partition prior knowledge in the fine-tuning of item similarities, bringing significant improvements in recommendation performance. Experimental results demonstrate the superiority of FPSR in efficiency, scalability and performance compared to state-of-the-art GCN models and item similarity models on CF tasks.

As for future work, we consider further exploring the role of item similarity models including FPSR in mitigating item popularity bias on CF task. Popularity bias of item has become popular in recent years on the studies of MF and GCN models. In contrast, in terms of item similarity models, it remains an open problem to be explored by researchers.
%%
%% The acknowledgments section is defined using the "acks" environment
%% (and NOT an unnumbered section). This ensures the proper
%% identification of the section in the article metadata, and the
%% consistent spelling of the heading.
\begin{acks}
This work was partially supported by the Research Grants Council of the Hong Kong Special Administrative Region, China (Project No. CityU 11216620), and the National Natural Science Foundation of China (Project No. 62202122).
\end{acks}

%%
%% The next two lines define the bibliography style to be used, and
%% the bibliography file.
\bibliographystyle{ACM-Reference-Format}
\bibliography{FPSR}

%%
%% If your work has an appendix, this is the place to put it.
\newpage
\appendix
\section{Pseudo-code For FPSR}
\begin{algorithm}[!]
\renewcommand{\algorithmicrequire}{\textbf{Input:}}
\renewcommand{\algorithmicensure}{\textbf{Output:}}
\renewcommand{\algorithmicloop}{\textbf{Function} \textsc{PART}($\mathcal{I}, R$)}
\renewcommand{\algorithmicendloop}{\algorithmicend\ \textbf{Function}}
\caption{Fine-tuning Partition-aware item Similarities for efficient and scalable Recommendation (FPSR)}\label{Alg:FPSR}
\begin{algorithmic}
\REQUIRE item set $\mathcal{I}$, interaction matrix $R$, hyperparameters $\lambda, \theta_1, \theta_2, \eta, \tau$
\ENSURE top-$K$ recommended items for user $u$
  \LOOP
  \STATE $V \gets TruncatedSVD(\tilde{R})$
  \STATE Derive $\mathcal{I}_1, \mathcal{I}_2$ from $\mathcal{I}$ through \eqref{Eq:Partitioning}
  \STATE $\mathcal{I}_P \gets EmptySet()$
  \FOR{$\mathcal{I}_n$ in $\{\mathcal{I}_1, \mathcal{I}_2\}$}
    \IF{$\lvert \mathcal{I}_n \rvert \geq \tau \lvert \mathcal{I} \rvert$}
        \STATE $\mathcal{I}_P \gets Merge(\mathcal{I}_P, \textit{PART}(\mathcal{I}_n, R_n))$
    \ELSE{}
        \STATE $\mathcal{I}_P \gets Merge(\mathcal{I}_P, \mathcal{I}_n)$
    \ENDIF
  \ENDFOR
  \STATE \textbf{return} $\mathcal{I}_P$ 
\ENDLOOP
\STATE $V \gets TruncatedSVD(\tilde{R})$
\STATE $W \gets D_I^{-\frac{1}{2}}VV^TD_I^{\frac{1}{2}}$
\STATE $\mathcal{I}_P \gets \textsc{PART} (\mathcal{I}, R)$
\STATE $S_{List} \gets EmptyList()$
\FOR{$\mathcal{I}_n$ in $\mathcal{I}_P$}
    \STATE $\hat{Q} \gets $ \eqref{Eq:Qhat}
    \FOR{$t=1;t \leq max\_iter;t++$}
        \STATE $Z_n,S_n,\Phi_n \gets$ \eqref{Eq:ADMM}
    \ENDFOR
    \STATE $S_{List} \gets Append(S_{List}, S_n)$
\ENDFOR
\STATE $S \gets $ \eqref{Eq:FullS}
\STATE $C \gets $ \eqref{Eq:C}
\STATE Recommend items for user $u$ by \eqref{Eq:Recommend}
\end{algorithmic}
\end{algorithm}

\section{Derivation of ADMM Optimization Problem}
Here, we show the full derivation of the optimization problem in Eq. \eqref{Eq:ADMM}. Consider the sub-problem in each partition of Eq. \eqref{Eq:Problem_ADMM},
\begin{equation}
\begin{aligned}
\underset{S_n}{\mathrm{argmin}}&\ \frac{1}{2}\lVert R_n - R_n(\lambda W_n+S_n)\rVert_F^2+\frac{\theta_2}{2}\lVert D_I(\lambda W_n+S_n) \rVert_F^2 \\
&+\theta_1 \lVert S_n \rVert_1 + \frac{\eta}{2} \lVert \mathbf{1}^T-\mathbf{1}^T(\lambda W_n+S_n) \rVert_F^2\\
s.t.&\ diag(S_n)=0,
\end{aligned}
\label{Eq:Problem_Partition}
\end{equation}
Then we have
\begin{equation}
\begin{aligned}
&\underset{S_n}{\mathrm{argmin}}\frac{1}{2}\lVert R_n - R_n(\lambda W_n+S_n)\rVert_F^2\\
=\ &\underset{S_n}{\mathrm{argmin}}\frac{1}{2}tr(S_n^TR_n^TR_nS_n)-tr((I-\lambda W_n)R_n^TR_nS_n).
\end{aligned}
\label{Eq:Sn_Form}
\end{equation}
Similarly, then the term $\lVert \mathbf{1}^T-\mathbf{1}^T(\lambda W_n+S_n) \rVert_F^2$ has the same form as Eq. \eqref{Eq:Sn_Form}. For the weighted $l_2$ regularization term, we have
\begin{equation}
\begin{aligned}
\ &\mathop{\arg\min}\limits_{S_n}\ \lVert D_I^{\frac{1}{2}}(\lambda W_n + S_n) \rVert_F^2\\
=\ &\mathop{\arg\min}\limits_{S_n}\ \lVert \lambda D_I^{\frac{1}{2}} W_n + D_I^{\frac{1}{2}}S_n -D_I \rVert_F^2\\
=\ &\mathop{\arg\min}\limits_{S_n}\ \lVert D_I^{\frac{1}{2}}-D_I^{\frac{1}{2}}(\lambda W_n + S_n) \rVert_F^2, \\
\end{aligned}
\end{equation}
such a derivation holds with the diagonal constraint $diag(S_n)=0$. As the weighted $l_2$ regularization term has the same form as Eq. \eqref{Eq:Sn_Form}, problem Eq. \eqref{Eq:Problem_Partition} can be converted to
\begin{equation}
\begin{aligned}
\underset{S_n}{\mathrm{argmin}}\ &\frac{1}{2}tr(S_n^T\hat{Q}S_n)-tr((I-\lambda W_n)\hat{Q}S_n)
+\theta_1 \lVert S_n \rVert_1\\
&s.t.\ diag(S_n)=0,
\end{aligned}
\end{equation}
where $\hat{Q}$ is $R_n^TR_n+\theta_2 D_I+\eta$. By assigning terms other than $l_1$ regularization to the dual variable $Z$ in ADMM, we can derive the optimization problem in Eq. \eqref{Eq:Problem_ADMM}.

\end{document}